\documentclass[aps,pra,twocolumn,superscriptaddress,longbibliography]{revtex4-2}

\usepackage{amsmath}
\usepackage{bm}
\usepackage{graphicx}
\usepackage[breaklinks=true,dvipdfmx]{hyperref}
\hypersetup{colorlinks=true,citecolor=magenta,linkcolor=magenta,urlcolor=magenta}
\usepackage{breakcites}

\begin{document}


\title{Controlling and exploring quantum systems by algebraic expression of adiabatic gauge potential}


\author{Takuya Hatomura}
\email[]{takuya.hatomura.ub@hco.ntt.co.jp}
\affiliation{NTT Basic Research Laboratories \& NTT Research Center for Theoretical Quantum Physics, NTT Corporation, Kanagawa 243-0198, Japan}

\author{Kazutaka Takahashi}
\affiliation{Institute of Innovative Research, Tokyo Institute of Technology, Kanagawa 226-8503, Japan}


\date{\today}

\begin{abstract}
Adiabatic gauge potential is the origin of nonadiabatic transitions. 
In counterdiabatic driving, which is a method of shortcuts to adiabaticity, adiabatic gauge potential can be used to realize identical dynamics to adiabatic time evolution without requiring slow change of parameters. 
We introduce an algebraic expression of adiabatic gauge potential. 
Then, we find that the explicit form of adiabatic gauge potential can be easily determined by some algebraic calculations. 
We demonstrate this method by using a single-spin system, a two-spin system, and the transverse Ising chain. 
Moreover, we derive a lower bound for fidelity to adiabatic time evolution based on the quantum speed limit. 
This bound enables us to know the worst case performance of approximate adiabatic gauge potential. 
We can also use this bound to find dominant terms in adiabatic gauge potential to suppress nonadiabatic transitions. 
We apply this bound to magnetization reversal of the two-spin system and to quantum annealing of the transverse Ising chain. 
Adiabatic gauge potential reflects structure of energy eigenstates, and thus we also discuss detection of quantum phase transitions by using adiabatic gauge potential. 
We find a signature of a quantum phase transition in the transverse Ising chain. 
\end{abstract}


\maketitle


%
%
\section{Introduction}

Quantum control is a fundamental element of quantum technologies~\cite{Acin2018}. 
To control a quantum system in a desired way, we change parameters of a Hamiltonian at each time. 
When we quickly change these parameters, nonadiabatic transitions, i.e., transitions between different energy eigenstates, take place. 
These nonadiabatic transitions are described by adiabatic gauge potential (AGP)~\cite{Kolodrubetz2017}. 
In contrast, when we change parameters slowly in time, a given system shows adiabatic time evolution, i.e., it tracks its energy eigenstate~\cite{Born1928,Kato1950}. 
Recently, quantum information processing based on adiabatic time evolution has been paid much attention~\cite{Albash2018}. 
An advantage of adiabatic time evolution as a control protocol is robustness with respect to various kinds of errors. 
On the other hand, a drawback is requirement of slow operation. 
Such long time processes are useless and suffer from decoherence.

Shortcuts to adiabaticity have been paid much attention as methods to speedup adiabatic time evolution~\cite{Demirplak2003,Berry2009,Chen2010,Torrontegui2013,Guery-Odelin2019}. 
Counterdiabatic (CD) driving is one of such methods~\cite{Demirplak2003,Berry2009}. 
In CD driving, we add AGP, or in other words, the CD term, to a reference Hamiltonian, of which we would like to realize adiabatic time evolution. 
AGP cancels out diabatic changes, and thus its dynamics is identical to adiabatic time evolution of a reference Hamiltonian even if its operation is conducted within short time. 
At the beginning, it was considered that construction of AGP requires knowledge of energy eigenstates of a given reference Hamiltonian and thus application range was limited~\cite{Demirplak2003,Berry2009,Muga2010,Chen2010a}, but soon it was extended to complex systems having special properties such as equivalence with two-level systems~\cite{DelCampo2012,Takahashi2013,Hatomura2018b}, scale invariance~\cite{DelCampo2013,Jarzynski2013,Deffner2014}, and classical nonlinear integrability~\cite{Okuyama2016}. 
It was also found that AGP for general systems can be constructed with the aid of the variational principle~\cite{Sels2017}. 
However, it is not clear whether or not we can always construct the exact AGP by using the variational approach.

Even if AGP is constructed, it usually includes non-local and many-body interactions, and thus it is often difficult to implement in experiments. 
Therefore, many ways to construct approximate AGP consisting of local and a few-body interactions have been proposed~\cite{Sels2017,Takahashi2013,Opatrny2014,Saberi2014,Campbell2015,Hatomura2017,Hatomura2018a,Ozguler2018}. 
Performance of approximate AGP has been evaluated by using various kinds of measures such as fidelity, residual energy, and so on. 
Numerical simulation of real time evolution is usually required to calculate these measures.

AGP is also useful to explore structure of quantum systems because it reflects properties of energy eigenstates. 
For example, we can use AGP as a probe for quantum chaos~\cite{Pandey2020}. 
It is also possible to find singular points of a quantum system in parameter space by studying optimal direction of AGP~\cite{Sugiura2020}.

In this paper, we introduce an algebraic expression of AGP. 
By rewriting a given reference Hamiltonian and its AGP in terms of Lie algebra, we find that the exact AGP can easily be constructed in a fully algebraic way. 
By using the relation to the variational approach, we can also construct approximate AGP in the algebraic way. 
Moreover, our approach reveals a condition for determining AGP, which is not clear in the formalism of the variational approach~\cite{Sels2017}. 
We also derive a lower bound for fidelity to adiabatic time evolution based on the quantum speed limit in a similar way to Ref.~\cite{Suzuki2020}. 
This bound can be used to evaluate the worst case performance of approximate AGP without simulating real time evolution. 
We can also use this bound to find dominant terms in AGP to suppress nonadiabatic transitions. 
Finally, we discuss detection of quantum phase transitions by using AGP. 
In particular, we focus on scaling behavior of AGP.

This paper is constructed as follows. 
In Sec.~\ref{Sec.algcd}, we introduce an algebraic expression of AGP and explain how to find the explicit form of it. 
This method is demonstrated by using a single-spin system. 
We also discuss relationship between the algebraic approach and the variational approach. 
In Sec.~\ref{Sec.evalu}, we derive a lower bound for fidelity to adiabatic time evolution. 
We apply this bound to magnetization reversal of a two-spin system to find dominant terms suppressing nonadiabatic transitions and to find the worst case performance when we use these terms as approximate AGP. 
We also apply it to quantum annealing of the transverse Ising chain. 
In Sec.~\ref{Sec.detect.qpt}, we discuss detection of quantum phase transitions by using AGP. 
Scaling behavior of approximate AGP for the transverse Ising chain around the critical point is found. 
Section~\ref{Sec.summary} is devoted to summary.

%
%
\section{\label{Sec.algcd}Algebraic expression of adiabatic gauge potential}

%
%
\subsection{\label{Sec.cd}Adiabatic gauge potential}

We consider a time-dependent Hamiltonian
\begin{equation}
\hat{H}(\bm{\lambda})=\sum_nE_n(\bm{\lambda})|n(\bm{\lambda})\rangle\langle n(\bm{\lambda})|,
\label{Eq.ham.eigen}
\end{equation}
where $E_n(\bm{\lambda})$ is the $n$th energy eigenvalue and $|n(\bm{\lambda})\rangle$ is the associated energy eigenstate. 
Here, $\bm{\lambda}=\bm{\lambda}_t$ is a set of time-dependent parameters. 
By moving to the adiabatic frame, where the Hamiltonian is diagonalized in the Schr\"odinger equation, we can find that nonadiabatic transitions are generated by AGP
\begin{equation}
\hat{\mathcal{A}}(\bm{\lambda})=i\sum_{n}(1-|n(\bm{\lambda})\rangle\langle n(\bm{\lambda})|)|\partial_{\bm{\lambda}}n(\bm{\lambda})\rangle\langle n(\bm{\lambda})|, 
\label{Eq.cdham.eigen}
\end{equation}
where $\hat{\mathcal{A}}(\bm{\lambda})$ is a vector of the Hermitian operators because of the differentiation $\partial_{\bm{\lambda}}=\partial/\partial\bm{\lambda}$~\cite{Kolodrubetz2017}. 
Throughout this paper, we set $\hbar=1$. 
In CD driving, we can realize adiabatic time evolution of the reference Hamiltonian (\ref{Eq.ham.eigen}) by adding the AGP (\ref{Eq.cdham.eigen}) as $\hat{H}(\bm{\lambda})+\dot{\bm{\lambda}}\cdot\hat{\mathcal{A}}({\bm{\lambda}})$~\cite{Demirplak2003,Berry2009}.

The AGP (\ref{Eq.cdham.eigen}) satisfies the condition~\cite{Kolodrubetz2017}
\begin{equation}
[\hat{H}(\bm{\lambda}),\partial_{\bm{\lambda}}\hat{H}(\bm{\lambda})-i[\hat{H}(\bm{\lambda}),\hat{\mathcal{A}}({\bm{\lambda}})]]=0.  
\label{Eq.cdham.cond}
\end{equation}
Note that if an Hermitian operator $\hat{A}({\bm{\lambda}})$ satisfies the condition (\ref{Eq.cdham.cond}) instead of the AGP (\ref{Eq.cdham.eigen}), it is also the AGP except for the diagonal part, which only affects phase factors. 
The variational approach is a method to search an Hermitian operator $\hat{A}({\bm{\lambda}})$ exactly or approximately satisfying the condition (\ref{Eq.cdham.cond}) based on the variational principle~\cite{Sels2017}. 
In this paper, we rather try to directly solve the condition (\ref{Eq.cdham.cond}).

%
%
\subsection{\label{Sec.sub.algcd}Algebraic construction}

We introduce the basis operators of $N$-dimensional Hilbert space $\{\hat{L}_\mu\}_{\mu=1,2,\dots,N^2-1}$ satisfying
\begin{equation}
\frac{1}{N}\mathrm{Tr}(\hat{L}_\mu\hat{L}_\nu)=\delta_{\mu\nu},
\label{Eq.Lie1}
\end{equation}
and
\begin{equation}
[\hat{L}_\mu,\hat{L}_\nu]=i\sum_\lambda f_{\mu\nu\lambda}\hat{L}_\lambda,
\label{Eq.Lie2}
\end{equation}
where $f_{\mu\nu\lambda}$ is an antisymmetric tensor. 
Here, we omit the identity operator. 
The reference Hamiltonian (\ref{Eq.ham.eigen}) and the AGP (\ref{Eq.cdham.eigen}) can be rewritten as
\begin{equation}
\hat{H}(\bm{\lambda})=\sum_{i=1}^Mh_{\mu_i}(\bm{\lambda})\hat{L}_{\mu_i},\quad\mu_i\in\{1,2,\dots,N^2-1\},
\label{Eq.ham.Lie}
\end{equation}
and
\begin{equation}
\hat{\mathcal{A}}({\bm{\lambda}})=\sum_{i=1}^{\tilde{M}}\alpha_{\tilde{\mu}_i}(\bm{\lambda})\hat{L}_{\tilde{\mu}_i},\quad\tilde{\mu}_i\in\{1,2,\dots,N^2-1\},
\label{Eq.cdham.Lie}
\end{equation}
where $\{h_{\mu_i}(\bm{\lambda})\}_{i=1,2,\dots,M}$ and $\{\alpha_{\tilde{\mu}_i}(\bm{\lambda})\}_{i=1,2,\dots,\tilde{M}}$ are sets of time-dependent parameters with integers $M,\tilde{M}\le N^2-1$. 
By multiplying $\hat{L}_{\tilde{\mu}_j}$ and taking the trace, we find the condition for the AGP (\ref{Eq.cdham.cond}) in the Lie algebraic expression
\begin{equation}
M(\bm{\lambda})\bm{\alpha}(\bm{\lambda})=\bm{u}(\bm{\lambda}), 
\label{Eq.alpha.vec}
\end{equation}
where $M(\bm{\lambda})$ is the $\tilde{M}\times\tilde{M}$ matrix with the matrix element 
\begin{equation}
M_{ij}(\bm{\lambda})=\mathrm{Tr}([\hat{H}(\bm{\lambda}),\hat{L}_{\tilde{\mu}_i}][\hat{H}(\bm{\lambda}),\hat{L}_{\tilde{\mu}_j}]), 
\end{equation}
$\bm{\alpha}(\bm{\lambda})={}^t(\alpha_{\tilde{\mu}_1}(\bm{\lambda}),\alpha_{\tilde{\mu}_2}(\bm{\lambda}),\dots,\alpha_{\tilde{\mu}_{\tilde{M}}}(\bm{\lambda}))$ is the $\tilde{M}$-dimensional vector, and $\bm{u}(\bm{\lambda})$ is the $\tilde{M}$-dimensional vector with the element 
\begin{equation}
u_i(\bm{\lambda})=i\mathrm{Tr}([\hat{H}(\bm{\lambda}),\partial_{\bm{\lambda}}\hat{H}(\bm{\lambda})]\hat{L}_{\tilde{\mu}_i}). 
\end{equation}
By solving Eq.~(\ref{Eq.alpha.vec}), we find the exact AGP (\ref{Eq.cdham.Lie}).

However, it is not always possible to directly solve Eq.~(\ref{Eq.alpha.vec}) due to the absence of the inverse matrix of $M(\bm{\lambda})$, which mathematically happens when the matrix $M(\bm{\lambda})$ is not full rank, i.e., $\mathrm{rank}(M(\bm{\lambda}))<\tilde{M}$. 
Physically, this reduction of the rank arises from symmetries of the reference Hamiltonian (\ref{Eq.ham.Lie}), which impose some constraints among $\{\alpha_{\tilde{\mu}_i}(\bm{\lambda})\}_{i=1,2,\dots,\tilde{M}}$, or arbitrariness in the diagonal part of the AGP (\ref{Eq.cdham.Lie}). 
In the former case, we can solve Eq.~(\ref{Eq.alpha.vec}) by reducing equivalent conditions in Eq.~(\ref{Eq.alpha.vec}). 
We will show this case in Sec.~\ref{Sec.ex.two} and \ref{Sec.ex.ising}. 
In the latter case, we can solve Eq.~(\ref{Eq.alpha.vec}) by adding another condition
\begin{equation}
\mathrm{Tr}\left(\hat{H}(\bm{\lambda})\hat{\mathcal{A}}(\bm{\lambda})\right)=0,
\label{Eq.another.cond}
\end{equation}
which fixes the diagonal part. 
We show this case in the next section.

Here we remark on the set of the basis operators for the AGP $\{\hat{L}_{\tilde{\mu}_i}\}_{i=1,2,\dots,\tilde{M}}$. 
By rewriting the AGP (\ref{Eq.cdham.Lie}) in the Lehmann's representation, we find that the AGP (\ref{Eq.cdham.Lie}) consists of the nested commutation relations
\begin{equation}
\hat{A}_k(\bm{\lambda})=[\hat{H}(\bm{\lambda}),[\hat{H}(\bm{\lambda}),\dots,[\hat{H}(\bm{\lambda}),\partial_{\bm{\lambda}}\hat{H}(\bm{\lambda})]\cdots]], 
\label{Eq.nested.comm}
\end{equation}
where the number of $\hat{H}(\bm{\lambda})$ in the nested commutation relation is $(2k-1)$ with an integer $k>0$~\cite{Claeys2019,Sugiura2020}. 
Namely, the AGP can be also expressed as
\begin{equation}
\hat{\mathcal{A}}(\bm{\lambda})=i\sum_{k=1}^\infty a_k(\bm{\lambda})\hat{A}_k(\bm{\lambda}), 
\label{Eq.cdham.nested}
\end{equation}
with appropriate parameters $\{a_k(\bm{\lambda})\}$~\cite{Claeys2019,Sugiura2020}. 
Therefore, we can find the set of the basis operators for the AGP $\{\hat{L}_{\tilde{\mu}_i}\}_{i=1,2,\dots,\tilde{M}}$ by calculating the odd nested commutation relations. 
We can easily confirm that the even nested commutation relations do not contribute to the AGP (\ref{Eq.cdham.Lie}) by calculating the off-diagonal elements of the condition (\ref{Eq.cdham.cond}). 
We also point out that to derive Eq.~(\ref{Eq.alpha.vec}), we focus on the $\tilde{M}$ components among the $(N^2-1)$ components by taking the trace with $\hat{L}_{\tilde{\mu}_i}$. 
This is justified by Eq.~(\ref{Eq.cdham.nested}) because it guarantees that both terms $[\hat{H}(\bm{\lambda}),\partial_{\bm{\lambda}}\hat{H}(\bm{\lambda})]$ and $[\hat{H}(\bm{\lambda}),[\hat{H}(\bm{\lambda}),\hat{\mathcal{A}}(\bm{\lambda})]]$ in the condition (\ref{Eq.cdham.cond}) only reproduce some of $\{\hat{L}_{\tilde{\mu}_i}\}_{i=1,2,\dots,\tilde{M}}$.

%
%
\subsection{Example}
Here we demonstrate our method by using a single-spin system
\begin{equation}
\hat{H}(\bm{\lambda})=h^x(\bm{\lambda})\hat{X}+h^y(\bm{\lambda})\hat{Y}+h^z(\bm{\lambda})\hat{Z},
\label{Eq.ham.single1}
\end{equation}
where we express the Pauli matrices as $\hat{X}$, $\hat{Y}$, and $\hat{Z}$. 
Note that the three basis operators of this two-dimensional system are given by $\{\hat{X},\hat{Y},\hat{Z}\}$. 
In particular, we show two situations. 
In the first situation, the AGP (\ref{Eq.cdham.Lie}) has no diagonal part, and thus Eq.~(\ref{Eq.alpha.vec}) can be directly solved. 
In contrast, in the second situation, the AGP (\ref{Eq.cdham.Lie}) has a diagonal part. 
Therefore, we solve Eq.~(\ref{Eq.alpha.vec}) by imposing the additional condition (\ref{Eq.another.cond}).

First, we consider the Hamiltonian (\ref{Eq.ham.single1}) with $h^y(\bm{\lambda})=0$. 
The nested commutation relations (\ref{Eq.nested.comm}) consist of only the basis operator $\hat{Y}$, and thus the AGP (\ref{Eq.cdham.Lie}) is given by
\begin{equation}
\hat{\mathcal{A}}(\bm{\lambda})=\alpha_Y(\bm{\lambda})\hat{Y},
\end{equation}
where the coefficient $\bm{\alpha}(\bm{\lambda})=\alpha_Y(\bm{\lambda})$ can be determined from Eq.~(\ref{Eq.alpha.vec}) and it is given by
\begin{equation}
\alpha_Y(\bm{\lambda})=\frac{h^z(\bm{\lambda})\partial_{\bm{\lambda}}{h}^x(\bm{\lambda})-h^x(\bm{\lambda})\partial_{\bm{\lambda}}{h}^z(\bm{\lambda})}{2[(h^{x}(\bm{\lambda}))^2+(h^{z}(\bm{\lambda}))^2]}. 
\end{equation}
This is of course identical to the known result constructed by using the energy eigenstates~\cite{Demirplak2003,Berry2009}.

Next, we consider the general case of the Hamiltonian (\ref{Eq.ham.single1}) with the nonzero coefficients $h^i(\bm{\lambda})\neq0$, $i=x,y,z$. 
Here, the nested commutation relations (\ref{Eq.nested.comm}) include all the basis operators $\hat{X}$, $\hat{Y}$, and $\hat{Z}$, and thus the AGP (\ref{Eq.cdham.Lie}) also consists of these basis operators
\begin{equation}
\hat{\mathcal{A}}(t)=\alpha_X(\bm{\lambda})\hat{X}+\alpha_Y(\bm{\lambda})\hat{Y}+\alpha_Z(\bm{\lambda})\hat{Z}. 
\label{Eq.cdham.single}
\end{equation} 
The time-dependent coefficients $\bm{\alpha}(\bm{\lambda})={}^t(\alpha_X(\bm{\lambda}),\alpha_Y(\bm{\lambda}),\alpha_Z(\bm{\lambda}))$ can be determined from Eq.~(\ref{Eq.alpha.vec}) with
\begin{equation}
M(\bm{\lambda})=-8
\begin{pmatrix}
h^{y2}+h^{z2} & -h^xh^y & -h^xh^z \\
-h^xh^y & h^{x2}+h^{z2} & -h^yh^z \\
-h^xh^z & -h^yh^z & h^{x2}+h^{y2}
\end{pmatrix}, 
\label{Eq.twolevel.m}
\end{equation}
and
\begin{equation}
\bm{u}(\bm{\lambda})=-4
\begin{pmatrix}
h^y\partial_{\bm{\lambda}}h^z-h^z\partial_{\bm{\lambda}}h^y \\
h^z\partial_{\bm{\lambda}}h^x-h^x\partial_{\bm{\lambda}}h^z \\
h^x\partial_{\bm{\lambda}}h^y-h^y\partial_{\bm{\lambda}}h^x
\end{pmatrix}, 
\label{Eq.malu.single1}
\end{equation}
where we abbreviate the parameter $\bm{\lambda}$ in the right-hand side of these equations for simplicity. 
However, the inverse matrix of the matrix (\ref{Eq.twolevel.m}) does not exist because the AGP (\ref{Eq.cdham.single}) has a diagonal part and the rank of the matrix (\ref{Eq.twolevel.m}) is two. 
To resolve this problem, we consider the additional condition (\ref{Eq.another.cond}), which gives
\begin{equation}
h^x(\bm{\lambda})\alpha_X(\bm{\lambda})+h^y(\bm{\lambda})\alpha_Y(\bm{\lambda})+h^z(\bm{\lambda})\alpha_Z(\bm{\lambda})=0. 
\label{Eq.another.cond.twolevel}
\end{equation}
Now we find that the left-hand side of Eq.~(\ref{Eq.alpha.vec}) is invariant under the transformation
\begin{equation}
M(\bm{\lambda})\to M(\bm{\lambda})-8
\begin{pmatrix}
h^{x2} & h^xh^y & h^xh^z \\
h^yh^x & h^{y2} & h^yh^z \\
h^zh^x & h^zh^y & h^{z2}
\end{pmatrix}, 
\end{equation}
because of the additional condition (\ref{Eq.another.cond.twolevel}), and then we can solve Eq.~(\ref{Eq.alpha.vec}). 
It is also identical to the known result constructed by using the energy eigenstates~\cite{Demirplak2003,Berry2009}.

Later, we will show other examples, but before that, we discuss relation between our method and the variational method.

%
%
\subsection{Relation to the variational approach}

In the variational approach, we introduce a trial AGP $\hat{\mathcal{A}}_\mathrm{tri}(\bm{\lambda})$ and the operator
\begin{equation}
\hat{G}(\bm{\lambda})=\partial_{\bm{\lambda}}\hat{H}(\bm{\lambda})-i[\hat{H}(\bm{\lambda}),\hat{\mathcal{A}}_\mathrm{tri}(\bm{\lambda})]. 
\label{Eq.tri.cd}
\end{equation}
Then, we construct the AGP based on the variational principle
\begin{equation}
\delta\|\hat{G}(\bm{\lambda})\|_\mathrm{HS}^2=0, 
\label{Eq.var.cd}
\end{equation}
with respect to the trial AGP $\hat{\mathcal{A}}_\mathrm{tri}(\bm{\lambda})$~\cite{Sels2017}. 
Here $\|\hat{O}\|_\mathrm{HS}\equiv\sqrt{\mathrm{Tr}(\hat{O}^2)}$ represents the Hilbert-Schmidt norm.  
We may find the exact AGP if the trial AGP $\hat{\mathcal{A}}_\mathrm{tri}(\bm{\lambda})$ includes all the possible operators. 
Notably, even if the trial AGP $\hat{\mathcal{A}}_\mathrm{tri}(\bm{\lambda})$ consists of only some limited operators, we can find an approximate AGP from Eq.~(\ref{Eq.var.cd}).

Now, we rewrite the trial AGP $\hat{\mathcal{A}}_\mathrm{tri}(\bm{\lambda})$ in terms of Lie algebra as
\begin{equation}
\hat{\mathcal{A}}_\mathrm{tri}(\bm{\lambda})=\sum_{i=1}^{\bar{M}}\bar{\alpha}_{\bar{\mu}_i}(\bm{\lambda})\hat{L}_{\bar{\mu}_i},\quad\bar{\mu}_i\in\{1,2,\dots,N^2-1\}, 
\label{Eq.tricdham.Lie}
\end{equation}
where $\{\bar{\alpha}_{\bar{\mu}_i}(\bm{\lambda})\}_{i=1,2,\dots,\bar{M}}$ is a set of time-dependent parameters with an integer $\bar{M}\le N^2-1$. 
The variational principle (\ref{Eq.var.cd}) in the Lie algebraic expression reads
\begin{equation}
\delta\|\hat{G}(\bm{\lambda})\|_\mathrm{HS}^2=\sum_{i=1}^{\bar{M}}\frac{\partial\|\hat{G}(\bm{\lambda})\|_\mathrm{HS}^2}{\partial\bar{\alpha}_{\bar{\mu}_i}(\bm{\lambda})}\delta\bar{\alpha}_{\bar{\mu}_i}(\bm{\lambda})=0, 
\end{equation}
and thus we find
\begin{equation}
\bar{M}(\bm{\lambda})\bar{\bm{\alpha}}(\bm{\lambda})=\bar{\bm{u}}(\bm{\lambda}),
\label{Eq.alpha.vec.2}
\end{equation}
where $\bar{M}(\bm{\lambda})$ is the $\bar{M}\times\bar{M}$ matrix with the matrix element
\begin{equation}
\bar{M}_{ij}(\bm{\lambda})=\mathrm{Tr}([\hat{H}(\bm{\lambda}),\hat{L}_{\bar{\mu}_i}][\hat{H}(\bm{\lambda}),\hat{L}_{\bar{\mu}_j}]),
\end{equation}
$\bar{\bm{\alpha}}(\bm{\lambda})={}^t(\bar{\alpha}_{\bar{\mu}_1}(\bm{\lambda}),\bar{\alpha}_{\bar{\mu}_2}(\bm{\lambda}),\cdots,\bar{\alpha}_{\bar{\mu}_{\bar{M}}}(\bm{\lambda}))$ is the $\bar{M}$-dimensional vector, and $\bar{\bm{u}}(t)$ is the $\bar{M}$-dimensional vector with the element 
\begin{equation}
\bar{u}_i(\bm{\lambda})=i\mathrm{Tr}([\hat{H}(\bm{\lambda}),\partial_{\bm{\lambda}}\hat{H}(\bm{\lambda})]\hat{L}_{\bar{\mu}_i}). 
\end{equation}
Therefore, if we consider the identical set of the basis operators to $\{\hat{L}_{\tilde{\mu}_i}\}_{i=1,2,\cdots,\tilde{M}}$, we can reproduce Eq.~(\ref{Eq.alpha.vec}). 
In this sense, the algebraic approach is equivalent to the variational approach.

In the algebraic approach, the condition for determining the exact AGP is clear. 
Namely, it can be found when the matrix $M(\bm{\lambda})$ is full rank. 
In contrast, in the variational approach, one cannot notice it until the variational operation (\ref{Eq.var.cd}) is performed. 
This is the most advantageous point of the algebraic approach compared with the variational approach. 
Moreover, the algebraic approach reveals that if we consider all the operators $\{\hat{L}_i\}_{i=1,2,\dots,N^2-1}$ for a trial AGP, we cannot find the exact AGP because it has a diagonal part, i.e., the rank of $M(\bm{\lambda})$ is less than $\tilde{M}$ and no inversion matrix exists.

%
%
\section{\label{Sec.evalu}Performance of counterdiabatic driving}

%
%
\subsection{\label{Sec.bound}Lower bound for fidelity}

In many cases, the exact AGP consists of many-body and non-local interactions, and thus it is difficult to implement in experiments. 
Therefore, we may use approximate AGP in CD driving. 
However, there is no guarantee that approximate AGP improves adiabaticity. 
Fidelity to adiabatic time evolution is one of the measures to evaluate performance of CD driving in the presence of approximations in AGP. 
In this section, we derive a lower bound for fidelity, which enables us to know performance of the worst case before simulating real time evolution. 
We point out that this bound can also be used to find dominant terms in the AGP to suppress nonadiabatic transitions.

We consider a total Hamiltonian
\begin{equation}
\hat{H}_\mathrm{tot}(t)=\hat{H}(\bm{\lambda}_t)+\dot{\bm{\lambda}}_t\cdot\hat{\mathcal{A}}_\mathrm{app}(\bm{\lambda}_t), 
\label{Eq.totham}
\end{equation}
where $\hat{\mathcal{A}}_\mathrm{app}(\bm{\lambda}_t)$ is approximate AGP for the reference Hamiltonian $\hat{H}(\bm{\lambda}_t)$. 
In this section, we explicitly express time dependence of the parameter $\bm{\lambda}=\bm{\lambda}_t$. 
Starting from an energy eigenstate $|n(\bm{\lambda}_0)\rangle$, fidelity to adiabatic time evolution is given by
\begin{equation}
p_n(t)=|\langle n(\bm{\lambda}_t)|\hat{U}(t)|n(\bm{\lambda}_0)\rangle|^2, 
\label{Eq.fid}
\end{equation}
where $\hat{U}(t)$ is the time evolution operator under the total Hamiltonian (\ref{Eq.totham}). 
Note that $p_n(t)=1$ when $\hat{\mathcal{A}}_\mathrm{app}(\bm{\lambda}_t)$ is the exact AGP.

Based on the quantum speed limit~\cite{Mandelstam1945,Margolus1998,Deffner2017}, we obtain a bound for the fidelity 
\begin{equation}
\arccos\sqrt{p_n(t)}\le\int_0^tdt^\prime\sigma[\dot{\bm{\lambda}}_{t^\prime}\cdot(\hat{\mathcal{A}}(\bm{\lambda}_{t^\prime})-\hat{\mathcal{A}}_\mathrm{app}(\bm{\lambda}_{t^\prime})),|n(\bm{\lambda}_{t^\prime})\rangle], 
\label{Eq.bound.qsl}
\end{equation}
where $\sigma[\hat{O},|\Psi\rangle]\equiv\sqrt{\langle\Psi|\hat{O}^2|\Psi\rangle-\langle\Psi|\hat{O}|\Psi\rangle^2}$ is the standard deviation, in a similar way to Ref.~\cite{Suzuki2020}. 
We can find that the right-hand side of this equation can be expressed as the line integral of $\bm{\lambda}_t$, and thus it does not depend on the total operation time but depends on the path of $\bm{\lambda}_t$~\cite{Suzuki2020,Rezakhani2010,Kolodrubetz2017,Hatomura2020}. 
When it is difficult to obtain the energy eigenstate $|n(\bm{\lambda}_t)\rangle$, we can replace the standard deviation in Eq.~(\ref{Eq.bound.qsl}) with the operator norm, $\|\hat{O}\|=\sqrt{\sup_{|\Psi\rangle}\langle\Psi|\hat{O}^2|\Psi\rangle}$, or the Hilbert-Schmidt norm although it is looser. 
Note that the bound (\ref{Eq.bound.qsl}) is a lower bound for Eq.~(\ref{Eq.fid}), and thus it guarantees the fidelity of the worst case. 
Indeed, by writing the bound as $\theta(t)\equiv\arccos\sqrt{p_n(t)}\le B(t)$, we can rewrite it as $p_n(t)=\cos^2\theta(t)\ge\cos^2B(t)$ for $B(t)\le\pi/2$. 
Remarkably, we can use this bound to find dominant terms in the exact AGP $\hat{\mathcal{A}}(\bm{\lambda}_t)$ to suppress nonadiabatic transitions from the integrand of the bound (\ref{Eq.bound.qsl}), which becomes large when dynamics tends to nonadiabatic, by considering each term in the exact AGP $\hat{\mathcal{A}}(\bm{\lambda}_t)$ as approximate AGP $\hat{\mathcal{A}}_\mathrm{app}(\bm{\lambda}_t)$. 
We will discuss this point in the next section.

%
%
\subsection{\label{Sec.ex.two}Example 1: Two-spin system}

\begin{figure}
\includegraphics[width=8cm]{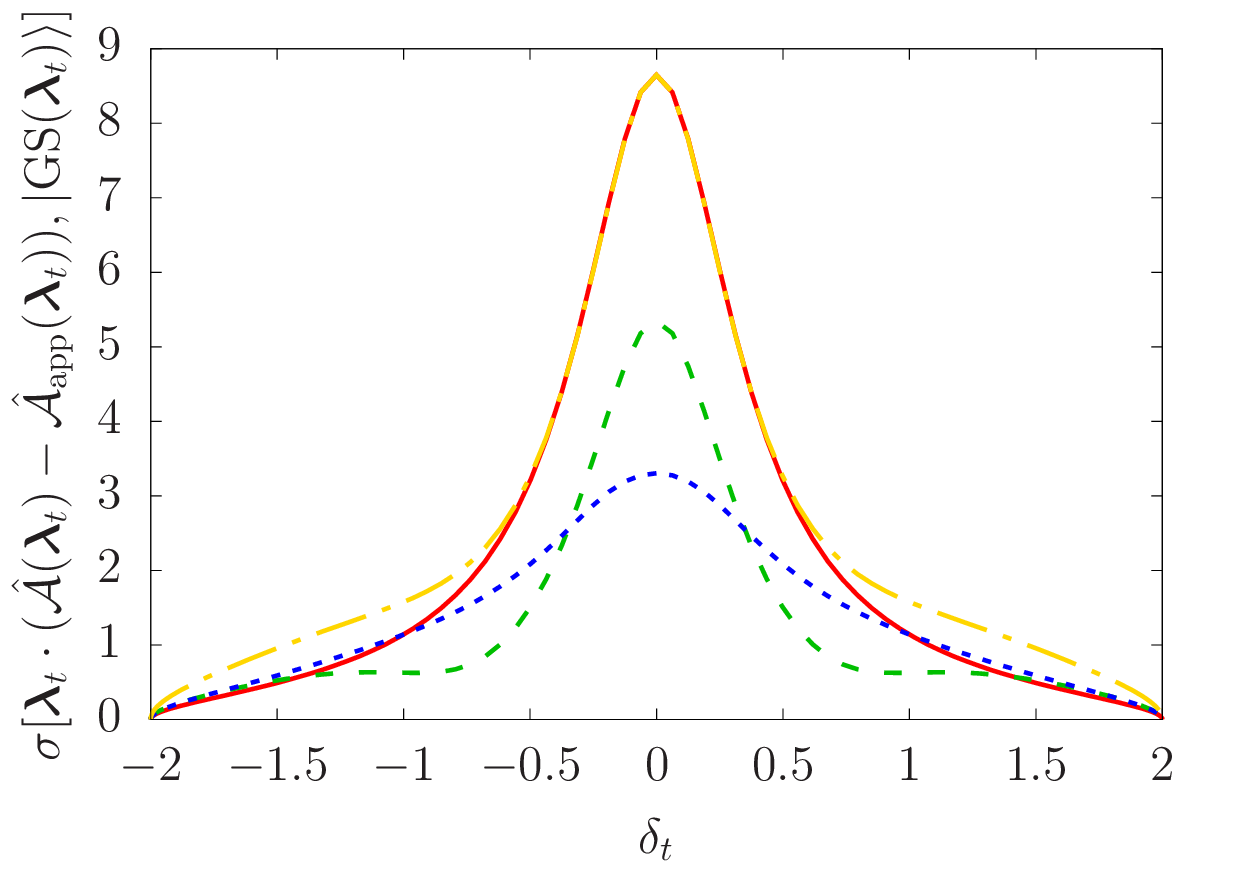}
\caption{\label{Fig.bd2spin} The integrand of the bound $\sigma[\dot{\bm{\lambda}}_t\cdot(\hat{\mathcal{A}}(\bm{\lambda}_t)-\hat{\mathcal{A}}_\mathrm{app}(\bm{\lambda}_t),|\mathrm{GS}(\bm{\lambda}_t)\rangle]$ with respect to the swept longitudinal field $\delta_t=-2\cos(\pi t/T)$ with $T=1$. The red solid curve is  (a) $\hat{\mathcal{A}}_\mathrm{app}(\bm{\lambda}_t)=0$ for reference, the green dashed curve is (b) $\hat{\mathcal{A}}_\mathrm{app}(\bm{\lambda}_t)=\alpha_Y(\bm{\lambda}_t)(\hat{Y}_1+\hat{Y}_2)$, the blue dotted curve is  (c) $\hat{\mathcal{A}}_\mathrm{app}(\bm{\lambda}_t)=\alpha_{XY}(\bm{\lambda}_t)(\hat{X}_1\hat{Y}_2+\hat{Y}_1\hat{X}_2)$, and the yellow dash-dotted curve is  (d) $\hat{\mathcal{A}}_\mathrm{app}(\bm{\lambda}_t)=\alpha_{YZ}(\bm{\lambda}_t)(\hat{Y}_1\hat{Z}_2+\hat{Z}_1\hat{Y}_2)$. }
\end{figure}

\begin{figure}
\includegraphics[width=8cm]{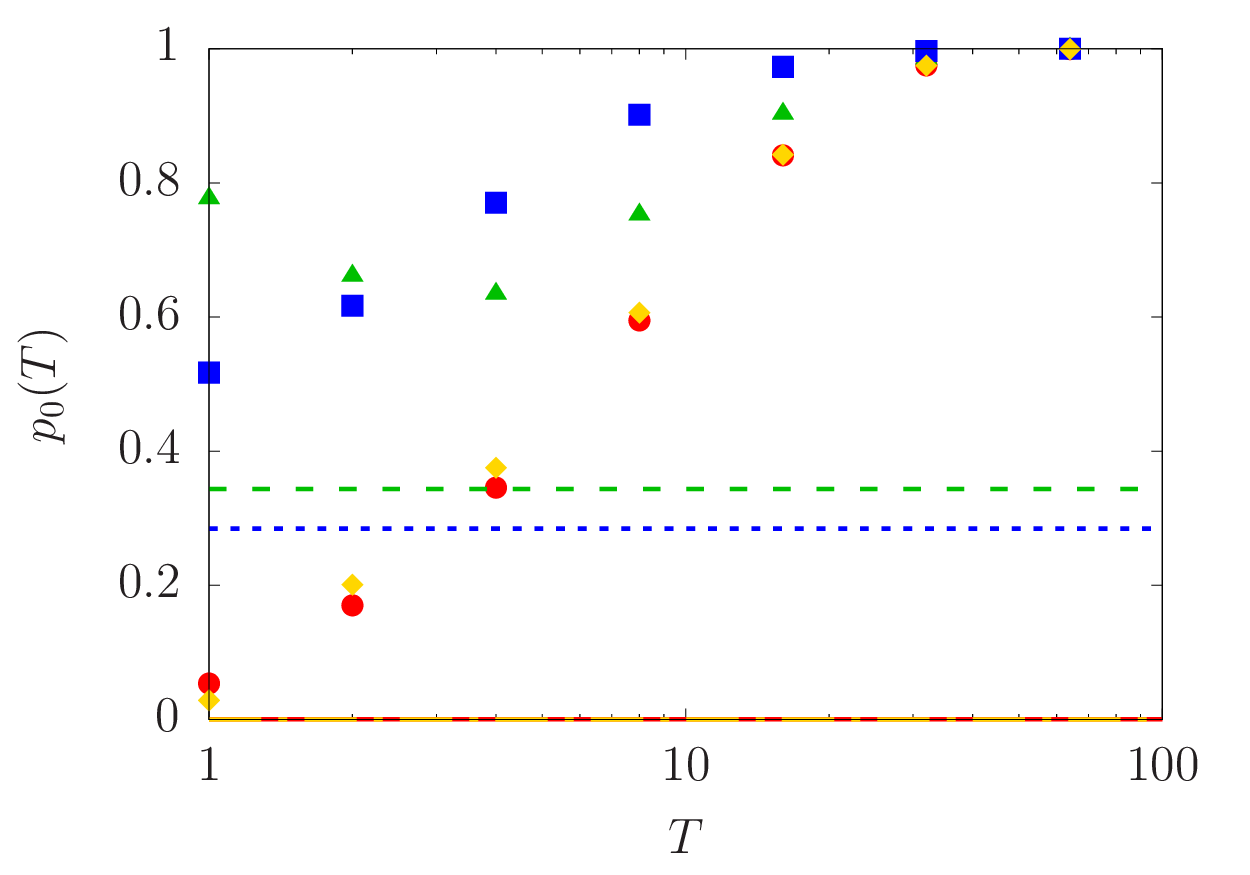}
\caption{\label{Fig.fidtime} The fidelity $p_0(T)=|\langle\mathrm{GS}(\bm{\lambda}_T)|\hat{U}(T)|\mathrm{GS}(\bm{\lambda}_0)\rangle|^2$ with respect to the operation time $T$. In a similar way to Fig.~\ref{Fig.bd2spin}, (red circles) the case (a), (green triangles) the case (b), (blue squares) the case (c), and (yellow diamonds) the case (d) are plotted with the bound (\ref{Eq.bound.qsl}), which is represented by the red solid line, the green dashed line, the blue dotted line, and the yellow dash-dotted line, respectively. }
\end{figure}

As the first example, we consider magnetization reversal of a two-spin system described by a Hamiltonian
\begin{equation}
\hat{H}(\bm{\lambda}_t)=\chi_0\hat{Z}_1\hat{Z}_2+\delta_t(\hat{Z}_1+\hat{Z}_2)+\Omega_0(\hat{X}_1+\hat{X}_2), 
\label{Eq.2spin.ham}
\end{equation}
where we write the Pauli matrices as $\hat{X}_i$, $\hat{Y}_i$, and $\hat{Z}_i$, ($i=1,2$), and $\bm{\lambda}_t=\{\chi_0,\delta_t,\Omega_0\}$ is a set of parameters. 
Here, we sweep the longitudinal field $\delta_t$ from negative to positive under the fixed parameters $\chi_0$ and $\Omega_0$. 
In this section, we construct the exact AGP for the reference Hamiltonian (\ref{Eq.2spin.ham}) based on the algebraic approach, and then we discuss the fidelity (\ref{Eq.fid}) and its lower bound (\ref{Eq.bound.qsl}) to find dominant terms to suppress nonadiabatic transitions.

The fifteen basis operators of this four-dimensional system are given by $\{\hat{X}_1$, $\hat{X}_2$, $\hat{Y}_1$, $\hat{Y}_2$, $\hat{Z}_1$, $\hat{Z}_2$, $\hat{X}_1\hat{X}_2$, $\hat{X}_1\hat{Y}_2$, $\hat{X}_1\hat{Z}_2$, $\hat{Y}_1\hat{X}_2$, $\hat{Y}_1\hat{Y}_2$, $\hat{Y}_1\hat{Z}_2$, $\hat{Z}_1\hat{X}_2$, $\hat{Z}_1\hat{Y}_2$, $\hat{Z}_1\hat{Z}_2\}$. 
The nested commutation relations (\ref{Eq.nested.comm}) produce the following operators: $\hat{Y}_1$, $\hat{Y}_2$, $\hat{X}_1\hat{Y}_2$, $\hat{Y}_1\hat{X}_2$, $\hat{Y}_1\hat{Z}_2$, and $\hat{Z}_1\hat{Y}_2$. 
Therefore, by taking into account the permutation symmetry of the reference Hamiltonian (\ref{Eq.2spin.ham}), the AGP is given by
\begin{equation}
\begin{aligned}
\hat{\mathcal{A}}(\bm{\lambda}_t)=&\alpha_Y(\bm{\lambda}_t)(\hat{Y}_1+\hat{Y}_2)+\alpha_{XY}(\bm{\lambda}_t)(\hat{X}_1\hat{Y}_2+\hat{Y}_1\hat{X}_2) \\
&+\alpha_{YZ}(\bm{\lambda}_t)(\hat{Y}_1\hat{Z}_2+\hat{Z}_1\hat{Y}_2), 
\end{aligned}
\label{Eq.cdham.twospin}
\end{equation}
where the coefficients $\bm{\alpha}(\bm{\lambda}_t)={}^t(\alpha_Y(\bm{\lambda}_t),\alpha_{XY}(\bm{\lambda}_t),\alpha_{YZ}(\bm{\lambda}_t))$ can be determined from Eq.~(\ref{Eq.alpha.vec}).

Now we discuss truncation of the exact AGP by assuming restriction of possible interactions and try to find dominant terms. 
We consider the following cases: (a) $\hat{\mathcal{A}}_\mathrm{app}(\bm{\lambda}_t)=0$ for reference, (b) $\hat{\mathcal{A}}_\mathrm{app}(\bm{\lambda}_t)=\alpha_Y(\bm{\lambda}_t)(\hat{Y}_1+\hat{Y}_2)$, (c) $\hat{\mathcal{A}}_\mathrm{app}(\bm{\lambda}_t)=\alpha_{XY}(\bm{\lambda}_t)(\hat{X}_1\hat{Y}_2+\hat{Y}_1\hat{X}_2)$, and (d) $\hat{\mathcal{A}}_\mathrm{app}(\bm{\lambda}_t)=\alpha_{YZ}(\bm{\lambda}_t)(\hat{Y}_1\hat{Z}_2+\hat{Z}_1\hat{Y}_2)$. 
For numerical simulation, we set $\chi_0=-1$, $\Omega_0=-1$, and $\delta_t=-2\cos(\pi t/T)$. 
Here, $T$ is the operation time. 
For these parameters, a small energy gap appears at $\delta_{T/2}=0$, and thus nonadiabatic transitions occur around this point. 
We plot the integrand of the bound (\ref{Eq.bound.qsl}) in Fig.~\ref{Fig.bd2spin} for $T=1$. 
It is clear that integrand of the bound (\ref{Eq.bound.qsl}) becomes large around $\delta_{T/2}=0$ reflecting occurrence of nonadiabatic transitions. 
Compared with the reference (a), we find that the cases (b) and (c) suppress the integrand of the bound (\ref{Eq.bound.qsl}), while the case (d) rather increases it. 
Therefore, we expect that the cases (b) and (c) improve adiabaticity.

Now we perform numerical simulation of real time evolution and confirm correctness of the above expectation.  
We plot the fidelity (\ref{Eq.fid}) with the bound (\ref{Eq.bound.qsl}) for various operation time $T$ in Fig.~\ref{Fig.fidtime}. 
We find that the cases (b) and (c) actually increase the fidelity compared with the reference (a), while the case (d) does not. 
Moreover, the bounds (\ref{Eq.bound.qsl}) for the cases (b) and (c) are finite, while these for the cases (a) and (d) are almost zero. 
Note that the bound (\ref{Eq.bound.qsl}) is constant as discussed in Sec.~\ref{Sec.bound}. 
This result indicates that for any operation time $T$ the fidelity remains finite if we apply $\alpha_Y(\bm{\lambda}_t)$ or $\alpha_{XY}(\bm{\lambda}_t)$.

One may be interested in why we do not use approximate construction (\ref{Eq.alpha.vec.2}), but truncate the exact AGP. 
In the approximate construction (\ref{Eq.alpha.vec.2}), we find $\bar{\alpha}_{XY}(\bm{\lambda}_t)=0$ and $\bar{\alpha}_{YZ}(\bm{\lambda}_t)=0$ because of $\bar{u}_{XY}(\bm{\lambda}_t)=0$ and $\bar{u}_{YZ}(\bm{\lambda}_t)=0$ for the present parameter set $\bm{\lambda}_t=\{\chi_0,\delta_t,\Omega_0\}$. 
This is why we consider truncation of the exact AGP.

%
%
\subsection{\label{Sec.ex.ising}Example 2: Transverse Ising chain}

\begin{figure}
\includegraphics[width=8cm]{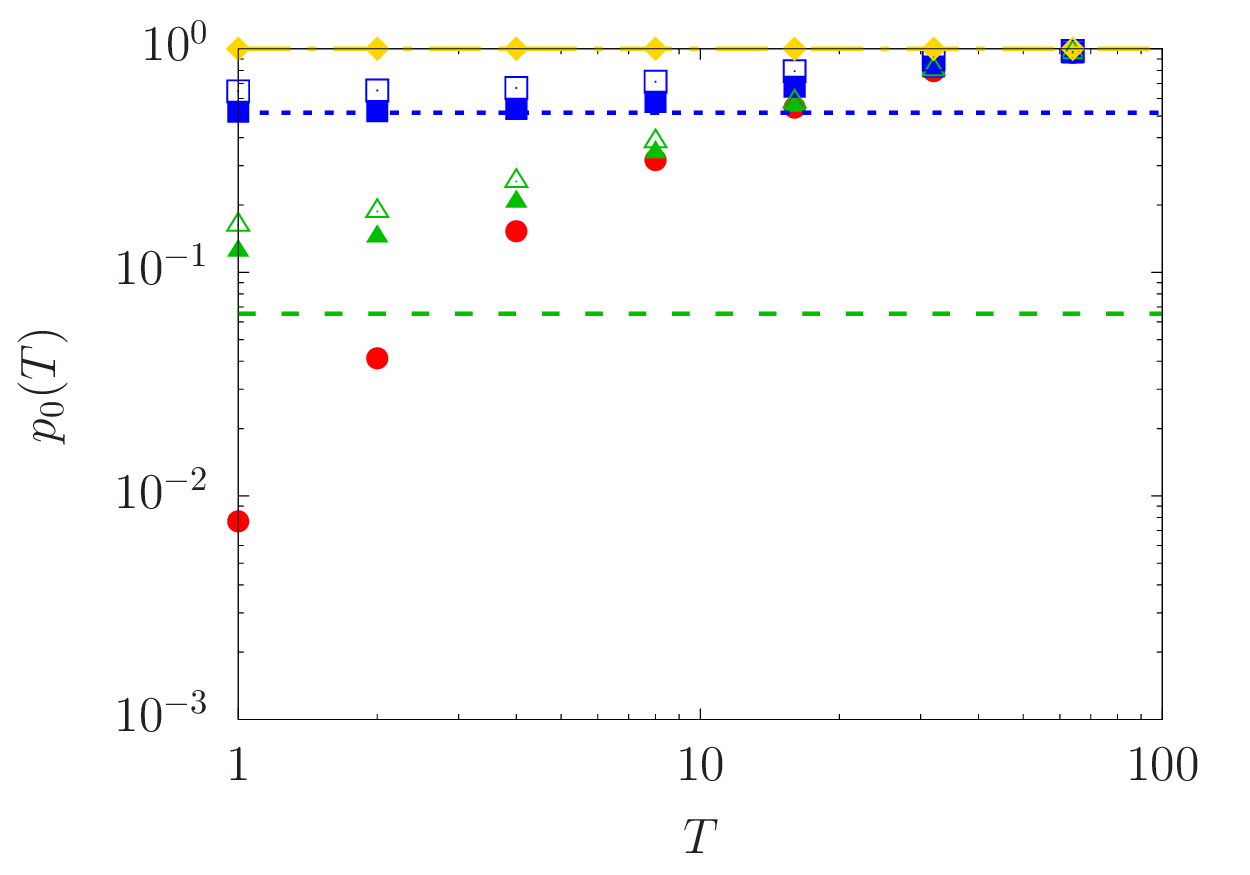}
\includegraphics[width=8cm]{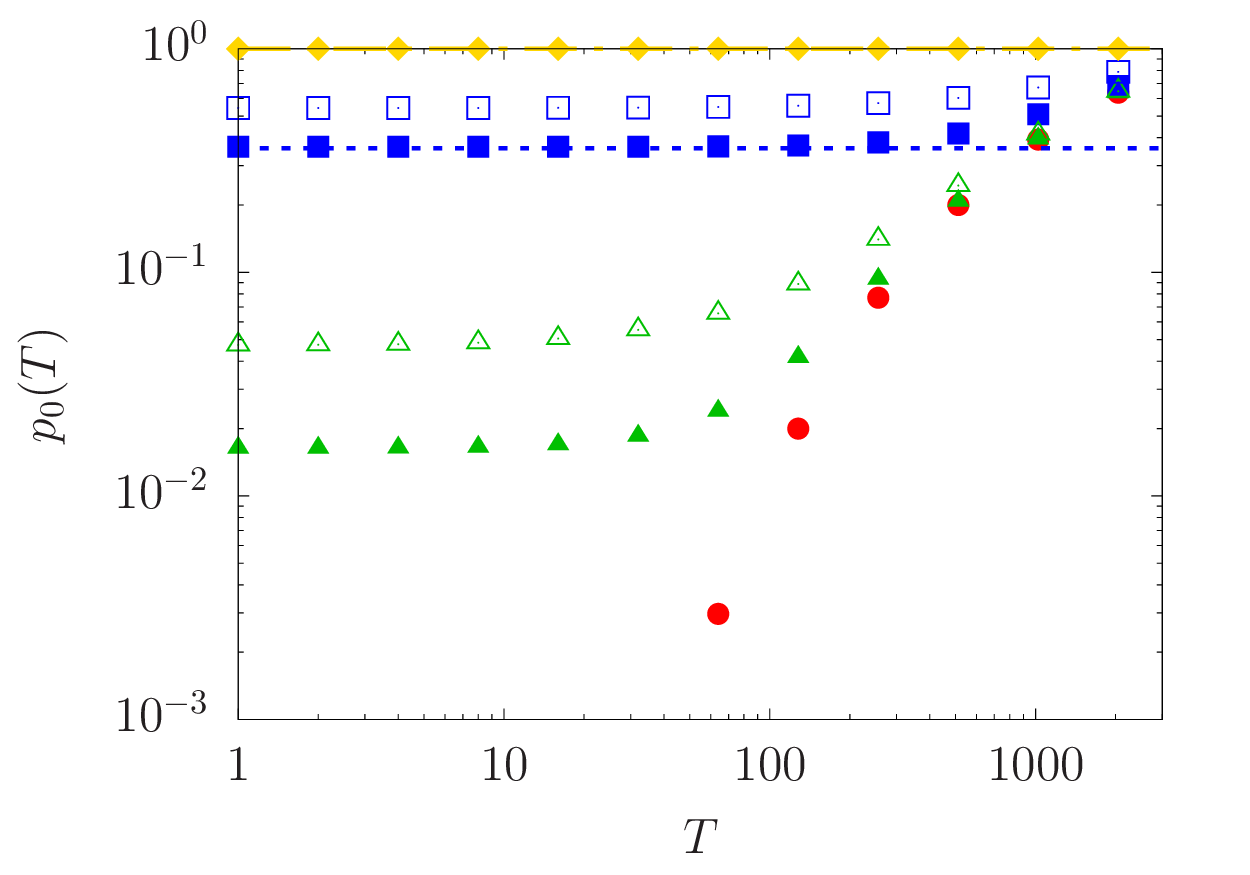}
\caption{\label{Fig.ising.fid} The fidelity $p_0(T)=|\langle\mathrm{GS}(\bm{\lambda}_T)|\hat{U}(T)|\mathrm{GS}(\bm{\lambda}_0)\rangle|^2$ with respect to the annealing time $T$. Here, the top panel is $L=10$ and the bottom panel is $L=100$. The open symbols represent (a) truncation of the exact AGP and the closed symbols represent (b) the approximate construction (\ref{Eq.alpha.vec.2}). The red circles are $\hat{\mathcal{A}}_\mathrm{app}(\bm{\lambda}_t)=0$ for reference, the green triangles are $K=0$ ($K=9$), the blue squares are $K=2$ ($K=29$), and the yellow diamonds are $K=8$ ($K=98$), for $L=10$ ($L=100$). For each data, we also plot the lower bound by the red solid line, the green dashed line, the blue dotted line, the yellow dash-dotted line, respectively. Note that some of the bounds are smaller than the bottom border. }
\end{figure}

Next, we consider quantum annealing of the Ising chain
\begin{equation}
\hat{H}(\bm{\lambda}_t)=-g_t\sum_{i=1}^L\hat{Z}_i\hat{Z}_{i+1}-(1-g_t)\sum_{i=1}^L\hat{X}_i,
\label{Eq.ham.ising}
\end{equation}
with the periodic boundary condition $\hat{Z}_{L+1}=\hat{Z}_1$. 
Here, the parameter $\bm{\lambda}_t$ is $g_t$ satisfying $g_0=0$ and $g_T=1$, where $T$ is the annealing time. 
In this model, the AGP is given by a series of many-body interactions~\cite{DelCampo2012}. 
In this section, we compare performance of two cases. 
One is truncation of the exact AGP~\cite{DelCampo2012} and the other is approximate construction of AGP (\ref{Eq.alpha.vec.2}).

The $4^{L}-1$ basis operators of this $2^L$-dimensional system can be generated by products of $\{\hat{X}_i,\hat{Y}_i,\hat{Z}_i\}_{i=1,2,\dots,L}$. 
The nested commutation relations (\ref{Eq.nested.comm}) produce the following operators: $\hat{Y}_i\hat{Z}_{i+1}$, $\hat{Z}_i\hat{Y}_{i+1}$, $\hat{Y}_i\hat{X}_{i+1}\hat{Z}_{i+2}$, $\hat{Z}_i\hat{X}_{i+1}\hat{Y}_{i+2}$, $\hat{Y}_i\hat{X}_{i+1}\hat{X}_{i+2}\hat{Z}_{i+3}$, $\hat{Z}_i\hat{X}_{i+1}\hat{X}_{i+2}\hat{Y}_{i+3}$, \dots, $\hat{Y}_i\hat{X}_{i+1}\hat{X}_{i+2}\cdots\hat{X}_{i+L-2}\hat{Z}_{i+L-1}$, and  $\hat{Z}_i\hat{X}_{i+1}\hat{X}_{i+2}\cdots\hat{X}_{i+L-2}\hat{Y}_{i+L-1}$ for all $i=1,2,\dots,L$. 
Therefore, by taking into account the translation symmetry of the reference Hamiltonian (\ref{Eq.ham.ising}), the AGP is given by
\begin{equation}
\begin{aligned}
\hat{\mathcal{A}}(\bm{\lambda}_t)=&\sum_{k=0}^{L-2}\alpha_{YXX\dots XZ}(\bm{\lambda}_t) \\
&\times\sum_{i=1}^L(\hat{Y}_i\hat{X}_{i+1}\hat{X}_{i+2}\cdots\hat{X}_{i+k}\hat{Z}_{i+k+1} \\
&\quad+\hat{Z}_i\hat{X}_{i+1}\hat{X}_{i+2}\cdots\hat{X}_{i+k}\hat{Y}_{i+k+1}), 
\end{aligned}
\label{Eq.AGP.Ising}
\end{equation}
where the number of $X$ in $\alpha_{YXX\dots XZ}(\bm{\lambda}_t)$ and the number of $\hat{X}$ between $\hat{Y}$ and $\hat{Z}$ are $k$. 
The coefficients $\alpha_{YXX\dots XZ}(\bm{\lambda}_t)$ can be easily obtained from Eq.~(\ref{Eq.alpha.vec}), which is $(L-1)$-dimensional linear equation. 
(a) For truncation of the exact AGP we restrict the summation in Eq.~(\ref{Eq.AGP.Ising}) up to $K$ ($K\le L-2$) and (b) for approximate construction (\ref{Eq.alpha.vec.2}) we use the identical set of the basis operators to the case (a).

We plot the fidelity (\ref{Eq.fid}) for various annealing time $T$ in Fig.~\ref{Fig.ising.fid}. 
We find that the case (a) shows slightly better results, but the case (b) also improve adiabaticity significantly. 
We also plot the bound (\ref{Eq.bound.qsl}) for the case (b) in Fig.~\ref{Fig.ising.fid}. 
The finite value of the bound (\ref{Eq.bound.qsl}) ensures that the fidelity never becomes zero for any annealing time $T$.

We stress that for many-body systems, in which large number of the basis operators appears in the AGP (\ref{Eq.cdham.Lie}), approximate construction (\ref{Eq.alpha.vec.2}) is much easier than truncation after the exact construction (\ref{Eq.alpha.vec}) because we can just take into account some low-order nested commutation relations (\ref{Eq.nested.comm}). 
Therefore, the similar behavior of the fidelity, which we can find in Fig.~\ref{Fig.ising.fid}, is attractive in practice. 
Indeed, for general chaotic systems, such as quantum annealing of spin glasses, we expect that exponentially large number of basis operators appears in the AGP (\ref{Eq.cdham.Lie}), and thus computational complexity to solve Eq.~(\ref{Eq.alpha.vec}) would increase in an exponential way. 
However, we could significantly improve adiabaticity with polynomial computational cost if a similar result to Fig.~\ref{Fig.ising.fid} also holds for quantum annealing of spin glasses.

%
%
\section{\label{Sec.detect.qpt}Detection of quantum phase transitions}

%
%
\subsection{Signatures of quantum phase transitions in adiabatic gauge potential}

The AGP reflects structure of energy eigenstates, and thus we expect that it can also be used to explore quantum systems themselves. 
Here, we discuss detection of quantum phase transitions by using the AGP. 
We consider the energy eigenstates with slightly different parameters, $|n(\bm{\lambda})\rangle$ and $|n(\bm{\lambda}+\delta\bm{\lambda})\rangle$. 
If these two quantum states are similar, the Fubini-Study distance between these two states, $d(|n(\bm{\lambda})\rangle,|n(\bm{\lambda}+\delta\bm{\lambda})\rangle)\equiv\arccos F(|n(\bm{\lambda})\rangle,|n(\bm{\lambda}+\delta\bm{\lambda})\rangle)$ with $F(|n(\bm{\lambda})\rangle,|n(\bm{\lambda}+\delta\bm{\lambda})\rangle)\equiv|\langle n(\bm{\lambda})|n(\bm{\lambda}+\delta\bm{\lambda})\rangle|$, can be expanded as~\cite{Zanardi2007}
\begin{equation}
\begin{aligned}
&[d(|n(\bm{\lambda})\rangle,|n(\bm{\lambda}+\delta\bm{\lambda})\rangle)]^2 \\
&\approx\sum_{i,j}\langle \partial_{\lambda_i}n(\bm{\lambda})|(1-|n(\bm{\lambda})\rangle\langle n(\bm{\lambda})|)|\partial_{\lambda_j}n(\bm{\lambda})\rangle\delta\lambda_i\delta\lambda_j. 
\end{aligned}
\label{Eq.FubiniStudy.eigen}
\end{equation}
This quantity is known as the quantum geometric tensor~\cite{Provost1980}. 
When a quantum phase transition happens within a certain parameter region $(\bm{\lambda},\bm{\lambda}+\delta\bm{\lambda})$, two quantum states $|n(\bm{\lambda})\rangle$ and $|n(\bm{\lambda}+\delta\bm{\lambda})\rangle$ should be significantly different, and thus we expect that the quantity (\ref{Eq.FubiniStudy.eigen}) shows singular behavior. 
Indeed, such singularity has been reported~\cite{Zanardi2006,Zanardi2007}.

It is known that the Hilbert-Schmidt norm of the AGP is given by the summation of the quantum geometric tensor~\cite{Funo2017}
\begin{equation}
\begin{aligned}
&\|\delta\bm{\lambda}\cdot\hat{\mathcal{A}}(\bm{\lambda})\|_\mathrm{HS}^2 \\
&=\sum_n\sum_{i,j}\langle \partial_{\lambda_i}n(\bm{\lambda})|(1-|n(\bm{\lambda})\rangle\langle n(\bm{\lambda})|)|\partial_{\lambda_j}n(\bm{\lambda})\rangle\delta\lambda_i\delta\lambda_j. 
\end{aligned}
\label{Eq.HSnorm.agp}
\end{equation}
Therefore, if the quantity (\ref{Eq.FubiniStudy.eigen}) shows singular behavior, the quantity (\ref{Eq.HSnorm.agp}) should also show similar singularity. 
Namely, we could find signatures of quantum phase transitions in Eq.~(\ref{Eq.HSnorm.agp}). 
Moreover, by using the property of the basis operators of Lie algebra (\ref{Eq.Lie1}), Eq.~(\ref{Eq.HSnorm.agp}) is rewritten as
\begin{equation}
\|\delta\bm{\lambda}\cdot\hat{\mathcal{A}}(\bm{\lambda})\|_\mathrm{HS}^2=N\sum_{i=1}^{\tilde{M}}[\delta\bm{\lambda}\cdot\alpha_{\tilde{\mu}_i}(\bm{\lambda})]^2. 
\label{Eq.agp.sum}
\end{equation}
This kind of expression was found in free systems~\cite{Pandey2020}, but it holds for any system if we consider the algebraic expression of the AGP (\ref{Eq.cdham.Lie}).

In this paper, we try to detect a quantum phase transition by using approximate AGP constructed by Eq.~(\ref{Eq.alpha.vec.2}) instead of the exact AGP. 
As mentioned in the previous section, it is much attractive if we could detect quantum phase transitions by using approximate AGP. 
Note that flow of the AGP was discussed in Ref.~\cite{Sugiura2020} to find singularity in systems including quantum phase transitions, but we rather discuss scaling behavior of the AGP because quantum phase transitions in principle make sense in the thermodynamic limit.

%
%
\subsection{Example}

\begin{figure}
\includegraphics[width=8cm]{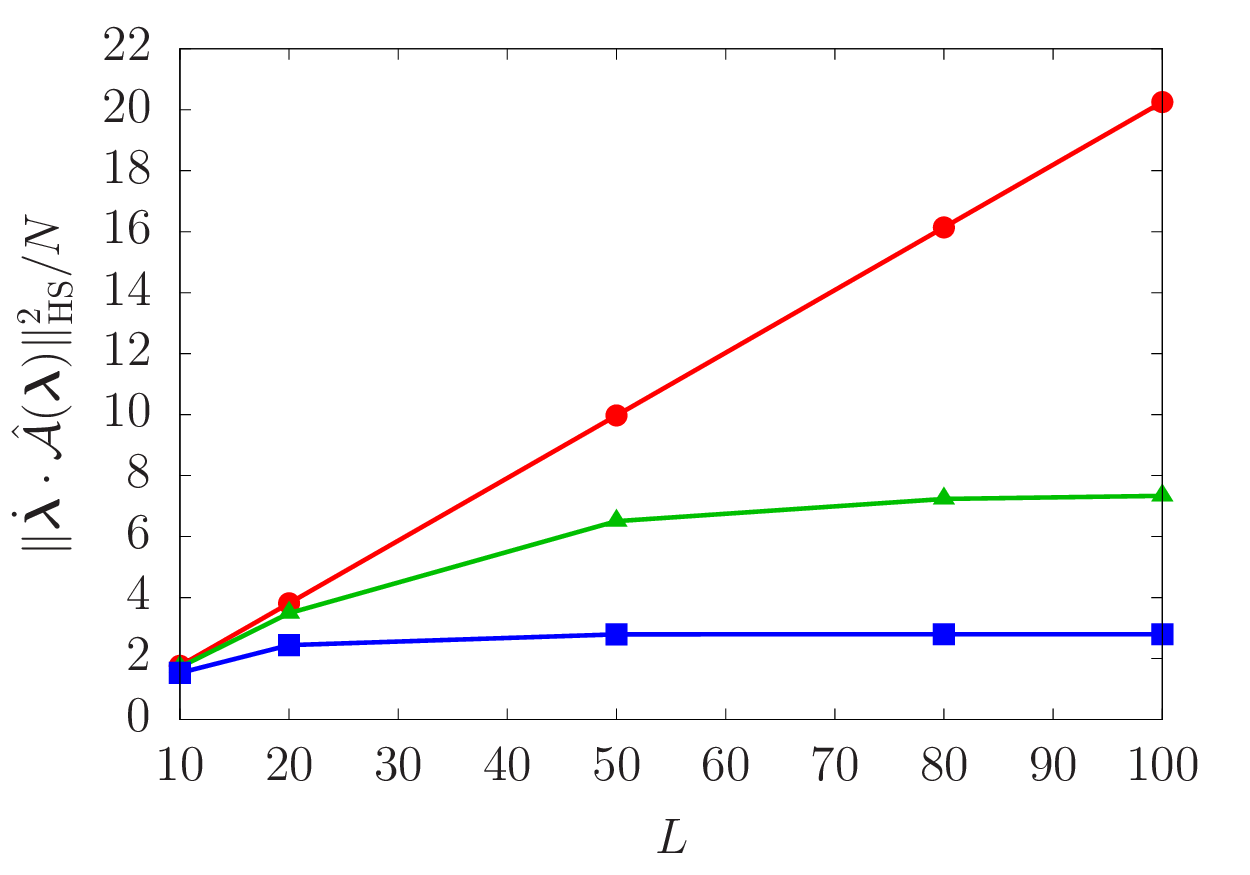}
\caption{\label{Fig.scale} The quantity $\|\delta\bm{\lambda}\cdot\hat{\mathcal{A}}(\bm{\lambda})\|_\mathrm{HS}^2/N$ with respect to the system size $L$. The red circles represent the critical point $g_t=0.5$, the green triangles represent $g_t=0.48$, and the blue squares represent $g_t=0.45$. }
\end{figure}

\begin{figure}
\includegraphics[width=8cm]{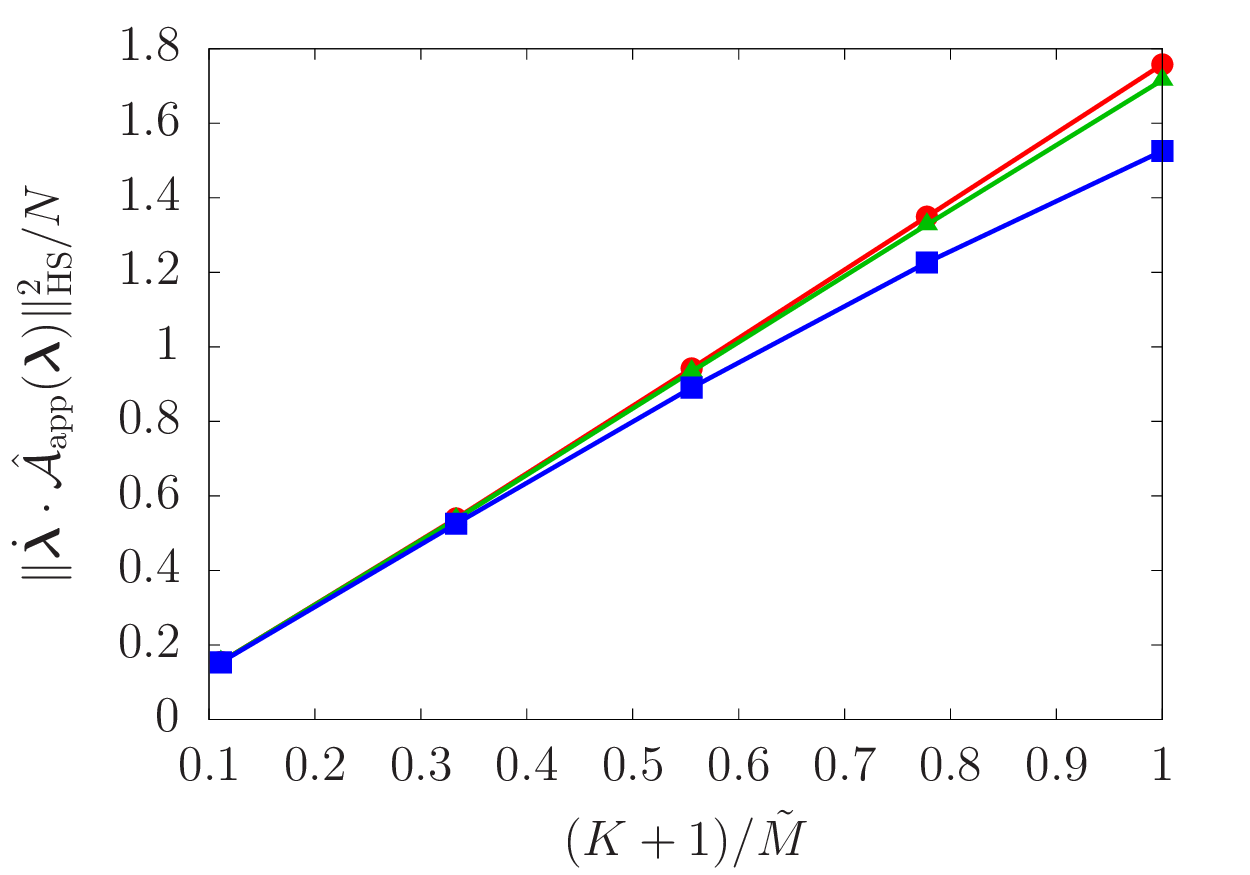}
\includegraphics[width=8cm]{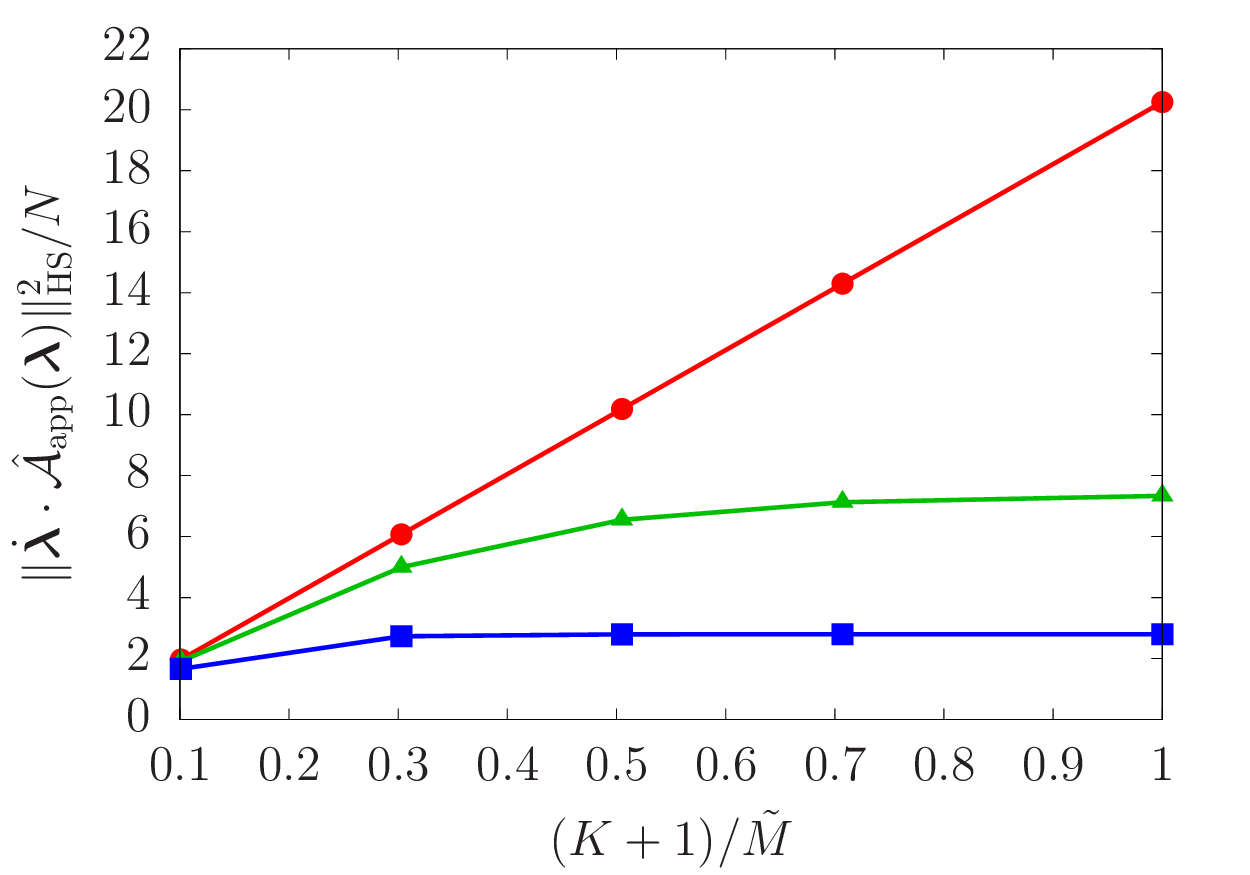}
\caption{\label{Fig.qgtpeak} The quantity $\|\delta\bm{\lambda}\cdot\hat{\mathcal{A}}_\mathrm{app}(\bm{\lambda})\|_\mathrm{HS}^2/N$ with respect to the restriction rate $(K+1)/\tilde{M}$ of the basis operators. The top panel is $L=10$ and the bottom panel is $L=100$. The red circles represent the critical point $g_t=0.5$, the green triangle represent $g_t=0.48$, and the blue squares represent $g_t=0.45$. }
\end{figure}

We again consider the transverse Ising chain (\ref{Eq.ham.ising}) with the periodic boundary condition. 
It is known that a quantum phase transition occurs at the critical point $g_t=1/2$, and thus we expect that a peak appears in Eq.~(\ref{Eq.agp.sum}) at there. 
Indeed, we can find it, but it is rather natural. 
We can find a peak even in the Landau-Zener process of a two-level system, whereas it is not a phase transition.  
Therefore, we focus on scaling behavior of Eq.~(\ref{Eq.agp.sum}).

First, we plot Eq.~(\ref{Eq.agp.sum}) with respect to the system size $L$ in Fig.~\ref{Fig.scale}. 
We find that it shows linear scaling at the critical point $g_t=1/2$, but it shows weaker behavior around there. 
Next, we consider approximate construction (\ref{Eq.alpha.vec.2}) for this method. 
Here, we restrict the basis operators up to $K$th order in the approximate construction (\ref{Eq.alpha.vec.2}) as we did in Sec.~\ref{Sec.ex.ising} and also in the summation in Eq.~(\ref{Eq.agp.sum}). 
We plot the restricted version of Eq.~(\ref{Eq.agp.sum}) with respect to the restriction rate $(K+1)/\tilde{M}$ in Fig.~\ref{Fig.qgtpeak}. 
We find that it also shows linear scaling at the critical point $g_t=1/2$, but it shows weaker behavior around there. 
We conclude that the linear scaling is a signature of the quantum phase transition in the transverse Ising chain.

%
%
\section{\label{Sec.summary}Summary}

In this paper, we introduced the algebraic expression of the AGP. 
The explicit form of both the exact and approximate AGP can be easily determined by the algebraic calculations. 
Although this algebraic approach is equivalent to the variational approach proposed in Ref.~\cite{Sels2017}, the algebraic approach clarifies the condition for obtaining the AGP. 
We also derived the lower bound for the fidelity to adiabatic time evolution based on the quantum speed limit. 
This bound gives the worst case performance of approximate AGP without numerical simulation of real time evolution. 
We can also use this bound to find dominant terms in the AGP to suppress nonadiabatic transitions. 
Finally, we discussed detection of quantum phase transitions by using approximate AGP.

We found the linear scaling behavior of the approximate AGP for the transverse Ising chain with respect to the restriction rate. 
We expect that other critical systems also show some scaling behavior, but it may not be linear. 
We leave study of approximate AGP for other systems as the future work.

\begin{acknowledgments}
KT was supported by JSPS KAKENHI Grant No. JP20K03781 and No. JP20H01827. 
\end{acknowledgments}

\bibliography{algcdbib}

\end{document}